**SUMMARY**

Natural open joints in rocks commonly present multi-scale self-affine apertures. This geometrical complexity affects fluid transport and heat exchange between the flowing fluid and the surrounding rock. In particular, long range correlations of self-affine apertures induce strong channeling of the flow which influences both mass and heat advection. A key question is to find a geometrical model of the complex aperture that describes at best the macroscopic properties (hydraulic conductivity, heat exchange) with the smallest number of parameters. Solving numerically the Stokes and heat equations with a lubrication approximation, we show that a low pass filtering of the aperture geometry provides efficient estimates of the effective hydraulic and thermal properties (apertures). A detailed study of the influence of the bandwidth of the lowpass filtering on these transport properties is also performed. For instance, keeping the information of amplitude only of the largest Fourier length scales allows us to reach already an accuracy of 9% on the hydraulic and the thermal apertures.

# Hydraulic transmissivity and heat exchange efficiency of open fractures: a model based on lowpass filtered apertures

Amélie Neuville[1,2,3], Renaud Toussaint[1,4] and Jean Schmittbuhl[1,5]
[1] *Institut de Physique du Globe de Strasbourg (IPGS), CNRS, and University of Strasbourg (EOST), 5 rue Descartes, 67000 trasbourg, France.*
[2] *Now at: Department of Physics, University of Oslo, P.O. Box 1048, Blindern, NO-0316 Oslo, Norway*
[3] *E-mail: amelie.neuville@unistra.fr*
[4] *E-mail: renaud.toussaint@unistra.fr*
[5] *E-mail: jean.schmittbuhl@unistra.fr*

19 June 2011

**SUMMARY**
Natural open joints in rocks commonly present multi-scale self-affine apertures. This geometrical complexity affects fluid transport and heat exchange between the flowing fluid and the surrounding rock. In particular, long range correlations of self-affine apertures induce strong channeling of the flow which influences both mass and heat advection. A key question is to find a geometrical model of the complex aperture that describes at best the macroscopic properties (hydraulic conductivity, heat exchange) with the smallest number of parameters. Solving numerically the Stokes and heat equations with a lubrication approximation, we show that a low pass filtering of the aperture geometry provides efficient estimates of the effective hydraulic and thermal properties (apertures). A detailed study of the influence of the bandwidth of the lowpass filtering on these transport properties is also performed. For instance, keeping the information of amplitude only of the largest Fourier length scales allows us to reach already an accuracy of 9% on the hydraulic and the thermal apertures.

## 1 INTRODUCTION

Mass and heat transport in natural fractures and joints are important in many contexts, notably in crystalline or carbonate rocks. Indeed, in addition to Darcian transport characteristic of many porous media (Bear 1972), the structure of some porous media involves a large range of scales and shapes (Dullien 1992), and networks of fractures sometimes carry a very significant part of the fluid in the rocks, affecting significantly the transport dynamics (Sahimi 1995a,b; Adler & Thovert 1999). Models of fluid transport in fractures of various shapes have been obtained analytically and numerically (Zimmerman & Bodvarsson 1996; Zimmerman & Main 2004; Méheust & Schmittbuhl 2001). The understanding of the transport in these units is of key importance to model oil transport in carbonate reservoirs, water or pollutant transport in calcareous or granitic rocks, as e.g. in water resource reservoirs (Plœmeur, France – Le Borgne et al. 2004) or around nuclear waste storage facilities in granitic rocks (Aspö, Sweden – Andersson et al. 2004) – see e.g. a recent review in Faybishenko et al. (2005).

For deep enhanced geothermal plants, which are a developing carbon-free energy producing technology, the geometry of the heat exchanger is also often dominated by a few large fracture zones. This is the case, for instance, in the granitic formation between the injection and extraction wells of the pilot power plant of Soultz-sous-Forêts, France (Dezayes et al. 2010), developed by the European project EHDRA (European Hot Dry Rock Association) over the last 30 years. Even in such a case, the heat exchanger does not consist of only few large fractures, but of interconnected large and small fractures regrouped in fracture zones (Genter & Traineau 1996; Sausse 2002). Such complexity can be described using an interconnected fracture network, as e.g. in the works of Bruel & Cacas (1992); Bruel (1995); Kohl & Hopkirk (1995); Kohl et al. (1995); Jupe et al. (1995); Hayashi et al. (1999); Kolditz & Clauser (1998); Watanabe & Tahahashi (1995). In such networks, most modelings propose simple models for single fracture units. We propose in the current study to focus on such a single element, and to see how its own geometrical complexity affects its heat and mass transport properties. We will describe a simple way to capture the effect of this complexity. Ideally, the present study could be used to parameterize the hydraulic and thermal properties of each element of a fracture network – combining deterministic and stochastic aspects to include at best the partial pieces of information on the network geometry and the geometric features of these units.

More specifically, the basic models of flow or heat exchange often consider that the joints have a simple geometry, like a volume between two parallel fracture surfaces. Self-affine geometry was however observed for most natural surfaces in fractures (Brown & Scholz 1985; Power & Durham 1997; Bouchaud 1997; Neuville et al. 2011b), faults (Power et al. 1987; Renard et al. 2006; Candela et al. 2009) or stylolites (Ebner et al. 2009a,b; Koehn et al. 2007; Schmittbuhl et al. 2004). Thus, more realistic models of fractures have been proposed, which often consist in one of the two follow-ing models: the void between two identical self-affine fracture surfaces, opened and shifted with respect to each other, or the void between two statistically independent self-affine surfaces. Both geometries have been observed, for example, in natural cases, like the joints present at depth in a marl formation analyzed in Draix (Neuville et al. 2011b,a).

Models already exist on how the self-affine characteristics of fracture surfaces affect on average the hydraulic and transport properties. For example, studies have been carried out on series of fractures with a certain aperture and roughness amplitude (Méheust & Schmittbuhl 2001; Neuville et al. 2010c).

For an open fracture, a key question to model its hydraulic transmissivity and the efficiency of the heat exchange is how to capture its geometric complexity with a small number of geometric parameters. In practice, some knowledge about the geometry of a fracture can be acquired by the study of core logs (e.g. Neuville et al. 2011b,a) at scales between micrometers and centimeters – some scaling hypothesis being required to extrapolate the large scale aperture. Unfortunately, such logs are rare. Generally, subsurface imaging methods, as surface seismic or electromagnetic methods, or localization of acoustic emissions during fluid injection, give some large scale information on the geometry of a particular fracture (e.g. Sausse et al. 2010). With such information at various scales, a central question is how to determine the transmissivity of a particular joint or fracture.

In this manuscript, we will show that both the thermal and hydraulic properties of transport in a self-affine fracture are largely determined by the large scale Fourier modes of its aperture.

## 2 EXISTING MODELS OF FLOW IN VARIABLE APERTURE FIELDS

For the simplest geometries of parallel plates, the stationary flow of Newtonian fluids is characterized by its hydraulic transmissivity described by the cubic law (Zimmerman & Main 2004; Zimmerman & Bodvarsson 1996), i.e., $j$, the fluid flux integrated across the aperture follows

$$j = -[H^3/(12\eta)]\nabla P, \qquad (1)$$

where $H$ is the distance between the plates, $\eta$ the fluid dynamic viscosity, $\nabla P$ the gradient of the deviation from hydrostatic pressure.

For apertures of any shape, at low speeds (in the Stokes regime), $j$ is proportional to $\nabla P$ (Zimmerman & Bodvarsson 1996). The transmissivity of the fracture is characterized by its hydraulic aperture $H$, defined as follows: for a given pressure drop, $H$ would be the distance between parallel plates leading to the observed flux, following Eq. (1). For rough fractures, the hydraulic aperture $H$ differs from the geometric aperture, $A = \langle a \rangle$, which is the spatial average of the aperture field $a(x, y)$, with $(x, y)$ Cartesian coordinates in the fracture midplane. It was shown (Beran 1968) that in general, the hydraulic aperture $H$ is bounded by $1/\langle a^{-3}\rangle \leq H^3 \leq \langle a^3 \rangle$, where $\langle . \rangle$ refers to space averaging. The lower bound corresponds to apertures varying in series along the flow, and the upper one to variations in parallel.

In order to avoid the explicit computations of the hydraulic flow in complex aperture fields, approximated computations of $H$ have been proposed, based on several moments of the aperture field distribution (Brown 1987; Méheust & Schmittbuhl 2001). Classical parameters are the average of the aperture field, $A$, and the standard deviation, $\sigma = \left(\langle a^2 \rangle - \langle a \rangle^2\right)^{1/2}$ (Brown 1987; Patir & Cheng 1978; Zimmerman & Bodvarsson 1996; Al-Yaarubi et al. 2005; Méheust & Schmittbuhl 2001). Several expressions have been obtained analytically and numerically, for various models of apertures presenting small fluctuations as e.g. sinusoidal or sawtooth profiles (Zimmerman & Bodvarsson 1996; Zimmerman & Main 2004), or in self-affine fractures (Méheust & Schmittbuhl 2001, 2003). Notably, for small aperture variations, $\sigma/A \ll 1$, it was shown that to second order in $\sigma/A$ (Méheust & Schmittbuhl 2003; Zimmerman & Main 2004), $H^3 = A^3[1 - (3/2)(\sigma/A)^2]$. The result at larger order depends on the type of correlations of the aperture field considered (Zimmerman & Main 2004; Méheust & Schmittbuhl 2001). Statistical characterizations of $H/A$ (its mean and dispersion) have been realized for a large number of self-affine apertures as functions of $\sigma/A$ (Neuville et al. 2010c).

Fluid transport in rough fractures is marked by channeling flow, as has been observed experimentally (Tsang 1984; Méheust & Schmittbuhl 2000; Plouraboué et al. 2000). It was shown that among all the Fourier modes of the aperture, only a few large scale modes play an important role for the hydraulic properties of self-affine fractures (Schmittbuhl et al. 2008). In this manuscript, we will study such filter-based approximation for the hydraulic transmissivity of fractures, and precisely quantify its precision depending on the number of modes kept.

Concerning the thermal exchange behavior in rough fractures, several models have been developed, notably in the context of geothermal reservoir modeling (Hayashi et al. 1999). Many models consider thermal and mass flux in networks of simple fractures (Kolditz & Clauser 1998; Kohl & Hopkirk 1995; Kohl et al. 1995; Jupe et al. 1995; Bruel 1995). Each fracture in these models is a simple object, e.g. with parallel plane walls. In such geometries, analytical solutions exist (Gringarten & Sauty 1975; Rodemann 1979; Heuer et al. 1991). As was however underlined by Hayashi et al. (1999); Kolditz & Clauser (1998), the complexity of mechanical effects as poro-elastic or thermo-elastic ones, modifying the geometry of the fractures is likely to influence the flow geometry and the heat coupling. The hydrothermal properties of complex fault zones have been studied with variations of porosity, permeability and thermal conductivity (Rosener et al. 2007). With realistic self-affine apertures, it was shown that the thermal coupling between the fluid and rock is affected by the induced channeling, for example in large fractures connecting the wells of Soultz-sous-Forêts (Neuville et al. 2010b).

To our knowledge, there has been no study so far of the specific influence of the different Fourier modes of the aperture on the thermal exchange efficiency.

## 3 APERTURE AND FLOW MODEL

In natural rocks, the surfaces of fractures and joints are often found to be self-affine, most of them with a Hurst exponent around 0.8 (Schmittbuhl et al. 1993, 1995; Bouchaud 1997).

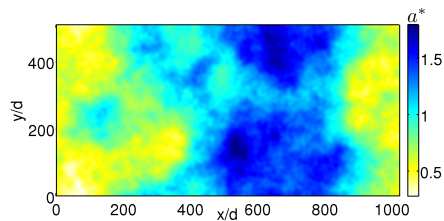

**Figure 1.** Self-affine synthetic aperture field with a normalized roughness amplitude $\sigma/A = 0.35$. The aperture field is normalized by its average, i.e. the map corresponds to $a(x,y)/A$. $d$ is a lattice step of arbitrary dimensions.

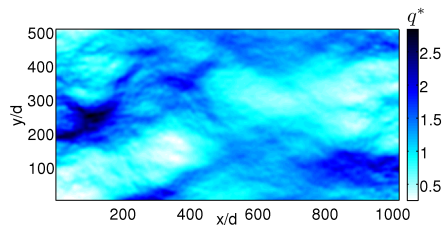

**Figure 2.** Integrated flow field $q/q_{\rm ref}$ in the aperture displayed: the color code is proportional to the norm of the flow.

If two such independent surfaces are facing each other, the aperture in-between will also be self-affine, with an identical Hurst exponent. Similarly, if a self-affine fracture surface is created, and the sides of the fracture are opened and sheared, the aperture will be self-affine at scales below the amount of shear (Plouraboué et al. 1995). The determination of the detailed morphology of the open volume in a fracture or a joint is difficult to perform *in situ*, and seldom done beyond the characterization of an average aperture. Recently, we have performed such a study on a natural example of rock (black marl) analyzed from a log core, and indeed found some example of self-affine aperture fields at the scale of observation (10 cm) (Neuville et al. 2011b,a).

The average aperture $A$ of such self-affine channels is not sufficient to characterize entirely their properties in terms of hydraulic transport and heat exchange efficiency between the rock and fluid (Neuville et al. 2010c; Schmittbuhl et al. 2008). Even if the additional knowledge of the aperture roughness amplitude $\sigma$ provides a better estimate of these properties, the dispersion between the different realizations of fractures with a given $(\sigma, A)$ is still large (Neuville et al. 2010c). We will see here how, for a specific self-affine fracture, the gradual degradation of the resolution with a lowpass spatial filter provides a good determination of the fluid mass transport and fluid-rock heat exchange properties.

Specifically, we will consider synthetic self-affine aperture fields $a(x, y)$ with a Hurst exponent of $\zeta = 0.8$, determined as follows (Méheust & Schmittbuhl 2001): on a grid of lattice step $d$, with $L_x \times L_y = 512d \times 1024d$ points regularly spaced in $(x, y)$, a white noise $\epsilon(x, y)$ is randomly generated. The 2D (two dimensional) Fourier transform of this noise $\tilde{\epsilon}(k_x, k_y)$ is evaluated, and multiplied by a scale dependent factor to obtain a varying field $\tilde{\psi}(k_x, k_y) = (k_x^2 + k_y^2)^{(-1-\zeta)/2}\tilde{\epsilon}(k_x, k_y)$ (Adler & Thovert 1999). The inverse Fourier transform provides a self-affine function $\psi(x, y)$ with the desired Hurst exponent. Normalizing it and adding it to a homogeneous value results in a self-affine aperture field of the chosen average aperture $A$ and standard deviation $\sigma$. Such an example of aperture, with $\sigma/A = 0.35$, is shown on Fig. 1. The flow considered is at low Reynolds number, i.e. in the Stokes regime, and the aperture variations are assumed small compared to the average aperture, so that a lubrication approximation allows to locally integrate the flux through the aperture thickness. The integrated flow $q = \int v(x, y, z)dz$ relates to the 2D pressure gradient

(e.g. Zimmerman & Bodvarsson 1996) as

$$q(x,y) = -\frac{a^3}{12\eta}\nabla P, \quad (2)$$

and it satisfies the 2D incompressibility equation: $\nabla \cdot q = 0$. The boundary conditions correspond to constant pressures imposed at $x = 0$ and at $x = L_x$, and impermeable sides parallel to the $x$ direction. After discretization on the grid, these equations are solved using a biconjugate gradient method (Press et al. 1992). The three-dimensional velocity $v(x, y, z)$ is related to $q(x, y)$ as

$$v(x,y,z) = q(x,y)[z_1(x,y) - z][z - z_2(x,y)]/a(x,y)^3, \quad (3)$$

where $z_1$ and $z_2$ are the out of plane coordinates of the fracture sides, related to the aperture by $z_2 - z_1 = a$. Since the flow only depends on the aperture, for simplicity, one considers here apertures symmetric around their mid planes, i.e. such as $z_2 = -z_1 = a/2$.

The obtained integrated flow $q$, for the aperture of Fig. 1, is shown on Fig. 2. The gray scale is proportional to $q^* = q/q_{\rm ref}$, where $q_{\rm ref}$ is the integrated flux between parallel plates spaced by the same average $A$, with the same imposed macroscopic gradient and viscosity. Note that the aperture dependence of the flow is complex: for instance, at coordinates $(600, 300)$, $q$ very low, while $a$ is quite large.

Profiles of the velocity along planes perpendicular to the fracture midplane, perpendicular or parallel to the average flow direction, are shown on Fig. 3(a) and (b). These velocities have been scaled by $v_{\rm refmax}$, the maximum velocity between parallel plates with the same geometric aperture. The heat transport is modeled by considering that in-plane advection entirely balances out-of-plane thermal conduction (Neuville et al. 2010c). Defining a thickness-averaged temperature weighted by the velocity,

$$\bar{T} = \int v(x,y,z)T(x,y,z)dz / \int v(x,y,z)dz, \quad (4)$$

these approximations lead to the following diffusion-advection equation for the thickness-averaged temperature

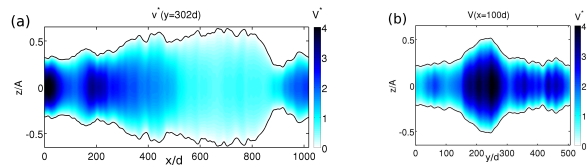

**Figure 3.** Cross section of the flow field $v(x, y, z)/v_{\rm refmax}$ across the aperture displayed, (a) parallel (at $y = 302d$) and (b) perpendicular (at $x = 100d$) to the flow direction.

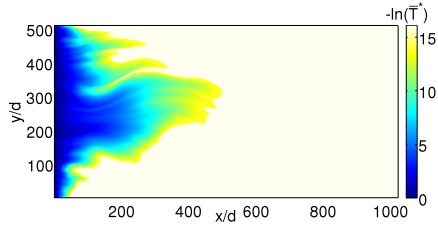

**Figure 4.** Integrated temperature field $-\ln(\bar{T}^*)$.

(Neuville et al. 2010c):
$$q \cdot \nabla \bar{T} + 2(\chi/a)\mathrm{Nu}(\bar{T} - T_r) = 0, \quad (5)$$

where $T_r$ is the rock temperature, $\chi$ is the thermal diffusivity of the fluid, and $\mathrm{Nu} = 70/17$ the Nusselt number in a Poiseuille flow for a fracture. This partial differential equation is also discretized by finite-differences and solved by a biconjugate gradient method (Neuville et al. 2010c). The temperature field obtained in this way for our example is shown on Fig. 4. The color scale corresponds to $-\ln(\bar{T}^*)$, with a dimensionless temperature:

$$\bar{T}^* = (\bar{T} - T_r)/(\bar{T}_{\mathrm{inlet}} - T_r), \quad (6)$$

where $\bar{T}_{\mathrm{inlet}}$ is the average temperature $\bar{T}(x=0)$ imposed at the inlet of the fracture, which corresponds to the injection of cold fluid. The three-dimensional temperature field, according to the thermal lubrication approximation, is

$$T(x,y,z) - T_r = (q \cdot \nabla \bar{T})/(2\chi a^3)(z-z_1)(z-z_2)(z-\sqrt{5}z_1)(z-\sqrt{5}z_2) \quad (7)$$

Cross sections of this temperature along the same planes as those used for the flow velocity are shown on Fig. 5(a) and (b) – with a dimensionless representation $T^* = (T - T_r)/(\bar{T}_{\mathrm{inlet}} - T_r)$.

## 4 FILTER BASED FLOW APPROXIMATION ON AN EXAMPLE

For any aperture field $a(x,y)$, with an integer $n$ called a filtering radius, the lowpass filtered field $a_n(x,y)$ is defined as follows from the 2D Fourier transform $\tilde{a}(k_x, k_y)$ of the aperture field:

$$\begin{cases} \left(\frac{k_x L_x}{2\pi}\right)^2 + \left(\frac{k_y L_y}{2\pi}\right)^2 \geq n^2 & \rightarrow \|\tilde{a}_n(k_x,k_y)^2\| = 0 \\ \left(\frac{k_x L_x}{2\pi}\right)^2 + \left(\frac{k_y L_y}{2\pi}\right)^2 < n^2 & \rightarrow \|\tilde{a}_n(k_x,k_y)^2\| = \|\tilde{a}(k_x,k_y)^2\| \end{cases}. \quad (8)$$

The real space aperture field $a_n(x,y)$ is obtained by inverse Fourier transform of $\tilde{a}_n$ so defined. For a discretized aperture of size $L_x \times L_y$, $(k_x, k_y)$ are wavevectors, and $(n_x, n_y) = (k_x L_x/(2\pi), k_y L_y/(2\pi))$ are integer wavenumbers. The filtering radius sets the number of large wavelength modes kept – see Fig. 6: for instance, for $n=1$, only the homogeneous aperture mode $(n_x, n_y) = (0,0)$ is kept in $a_1 = A$. For $n = 2$, the modes $(n_x, n_y) \in \{(0,0); (0,1); (1,0); (1,1); (-1,1)\}$ are kept (together with the complex conjugate modes, i.e. the opposite wavenumbers, (0,-1), (-1,0), (-1,-1), (1,-1)). We consider here the example of the self-affine fracture shown on Fig. 1. Applying the proposed lowpass filter of order $n$ generates apertures which are self-affine down to the scale $L_y/n$ (or for any shape $L_x$ and $L_y$, down to the scale $\min(L_x, L_y)/n$). The $n = 2$ and $n = 4$ filtered aperture fields are shown on Fig. 7.

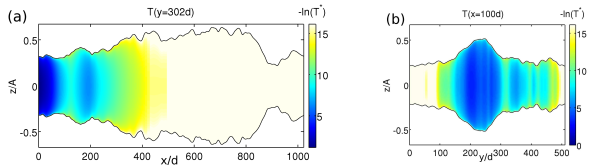

**Figure 5.** Cross section of the temperature field $-\ln(T^*)$, (a) parallel (at $y = 302d$) and (b) perpendicular to to the flow direction (at $x = 100d$).

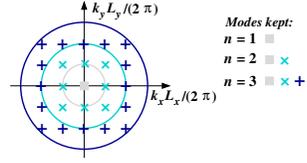

**Figure 6.** Representation of the lowpass spatial filter applied for filtering radii $n = 1$, $n = 2$ and $n = 3$: wavenumbers $(n_x, n_y)$ of the Fourier modes kept.

### 4.1 Hydraulic behavior

The same finite-difference numerical scheme as described previously is used to compute the flow field in these filtered apertures, according to Eqs. (2,3). The resulting flow velocity, integrated through the aperture is shown on Fig. 8. Two cross sections of the velocity field across the thickness are also computed and shown on Fig. 9. The comparison of these flow field structures to the results with the full resolution of the aperture roughness (Figs. 2 and 3) shows that the main features of the flow field are obtained from the computation of the flow in the filtered aperture. Thus, solving for the flow field in a lowpass filtered aperture seems to reproduce the main features of the flow. In order to evaluate more quantitatively the quality of this approximation, the hydraulic aperture $H_n$ is evaluated for the filtered apertures, as a function of the filtering radius $n$. The relative difference between $H_n$ and the hydraulic aperture with full resolution of the geometric aperture, $H_{\mathrm{ref}}$, is shown on Fig. 10(a). Clearly, this curve converge quickly. The reference aperture for this joint is such as $H_{\mathrm{ref}}/A = 0.816$, and this limit as $n$ grows is quickly approached by $H_n/A$. The trivial case of parallel plates corresponds to the $n = 1$ filtered case, for which $H/A = 1$ by definition, and for which $(H_1 - H_{\mathrm{ref}})/H_{\mathrm{ref}} = 22\%$. Already, including the four main modes in addition to the average aperture ($n = 2$) brings this relative accuracy to 0.8%: a very clear improvement, despite the small number of modes considered. In this example, all hydraulic apertures computed with radii $n > 6$ are closer than 0.2% from the full resolution $H_{\rm ref}$. The tendency observed in this example, that $H_n$ is overestimated at small $n$ with respect to $H_{\rm ref}$, is not systematic for all apertures: for some other cases, $H_{\rm ref} > A$, and $H_n$ is rather underestimated.

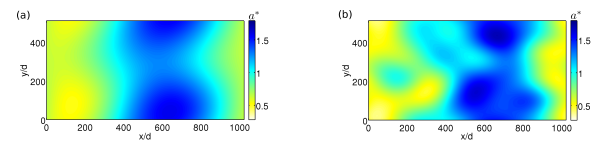

**Figure 7.** (a) Filtered aperture field $a/A$ with the 3 main Fourier modes and conjugates, i.e. a filtering radius $n = 2$, and (b) with a filtering radius $n = 4$.

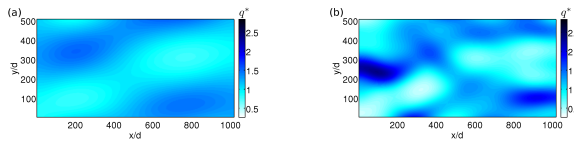

**Figure 8.** Dimensionless hydraulic flux $q/q_{\rm ref}$ in the filtered apertures, with filtering radii of respectively (a) $n = 2$ and (b) $n = 4$.

### 4.2 Thermal behavior

Similarly to the approach taken for the velocity, we compute the temperature field in large scale filtered apertures. The resulting thickness-averaged temperature field $\bar{T}$ is represented on Fig. (11). The corresponding cross sections are represented on Fig. (12). The scaling factors and the shading scales are identical to those used on Figs. (4) and (5). The similarities with these figures show how the large scales control the channeling of the energy transport to a great extent. In order to quantify the quality of the filtering approximation, a macroscopic quantity, named as the thermal aperture $\Gamma_n$, is evaluated in the apertures filtered at order $n$. At a coarse-grained scale, the temperature gap between the fluid and the rock evolves exponentially as a function of the travel distance in the channel. The inverse of the slope of the linear fit in a semilogarithmic representation of the temperature gap provides us a characteristic thermalization distance (Neuville et al. 2010c). This length characterizes the exponential evolution of the fluid temperature toward the rock one. The thermal aperture is then defined as the aperture between two parallel plates leading to the same characteristic thermalization distance (Neuville et al. 2010c). This characterizes the efficiency of the thermal exchange between the fluid and the surrounding rock. As an approximation to the thermal aperture of the rough aperture at full resolution, $\Gamma_{\rm ref}$, we consider the thermal aperture of the filtered aperture fields $\Gamma_n$. The relative difference between $\Gamma_n$ and the hydraulic aperture with full resolution of the geometric aperture, $\Gamma_{\rm ref}$, is shown on Fig. 10(b). Here, $\Gamma_{\rm ref}/A = 0.886$, i.e. the thermal aperture of the rough joint is $(\Gamma_1 - \Gamma_{\rm ref})/\Gamma_{\rm ref} = 11.4\%$ below the parallel plate case ($n = 1$). Considering the first largest Fourier modes ($n = 2$), the relative error of the approximation drops to 8%, and with the next largest wavelength ($n = 3$), below 4%. The speed of convergence of the filter-based approximation is slightly slower than for the hydraulic aperture. For $n \geq 6$, the precision of the approximation is better than 0.11%.

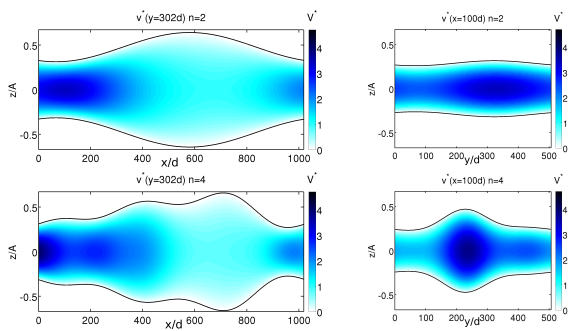

**Figure 9.** Cross section of the flow $v/v_{\rm refmax}$ in the $n = 2$ (top) and $n = 4$ (bottom) filtered aperture, (left) along the flow, at $y = 302d$ and (right) perpendicularly to the flow, at $x = 100d$.

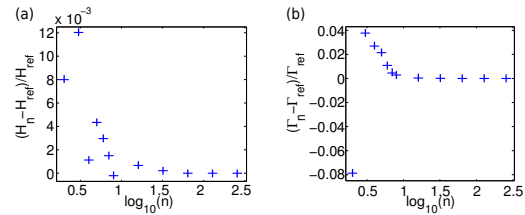

**Figure 10.** (a) Hydraulic and (b) thermal apertures as a function of the filter degree n (for $n \in \{2-8, 16, 32, 64, 128, 256\}$) relative to the fully rough case.

## 5 PRECISION OF THE FILTER-BASED APPROXIMATION, STATISTICS OVER THE APERTURES

The previous analysis of the accuracy of the filter approximation was for a particular aperture. We will study statistically the convergence of the hydraulic and thermal apertures as a function of the filter radius $n$ over many realizations. For this purpose, 300 realizations of statistically independent synthetic self-affine aperture (with the same Hurst exponent 0.8) are created. Each realization is characterized by a roughness amplitude $\sigma$ and an average aperture $A$. These 300 realizations are grouped in five values of $\sigma/A$, $\{0.1, 0.2, 0.3, 0.4, 0.5\}$, with at least 50 realizations in each group. The filtered hydraulic and thermal apertures are computed with filtering radii from $n = 1$ (flat) to $n = 11$ (around 100 largest scale Fourier modes) – see Fig. 13.

For each of the filtered morphologies, the flux and temperature are solved, the hydraulic aperture $H_n$ and thermal aperture $\Gamma_n$ are extracted, and compared to their counterparts for the unfiltered fields, $H_{\rm ref}$ and $\Gamma_{\rm ref}$. The roughness amplitude of the filtered aperture, $\sigma_n$, is also compared to that of the unfiltered aperture, $\sigma$. The resulting values as functions of the filter radius $n$ are shown on Fig. 13, for the group with the largest roughness, $\sigma/A = 0.5$. For these examples, the maximum relative error on the hydraulic aperture is 120% for $n = 1$ (flat plates), and 20% for $n = 2$ (4 largest modes). On the thermal one, the maximum relative error is 60% for $n = 1$, and 20% for $n = 2$. Obviously, adding even only the largest modes of the roughness increases sensibly the resolution of the hydraulic and thermal properties of the fracture model. The error on the amplitude of the roughness due to the filtering is at maximum 13% for $n \geq 2$. The value for $n = 1$ is not displayed, being by construction always equal to $(\sigma_1 - \sigma_{\text{ref}})/\sigma_{\text{ref}} = -1$.

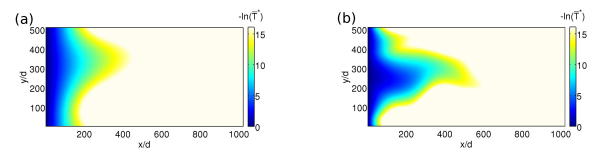

**Figure 11.** Temperature (integrated across the thickness) in a filtered aperture field, with filtering radii of respectively (a) $n = 2$ and (b) $n = 4$. The color scale represents $-\ln(\bar{T}^*)$, the logarithm of the scaled temperature $\bar{T}^*$.

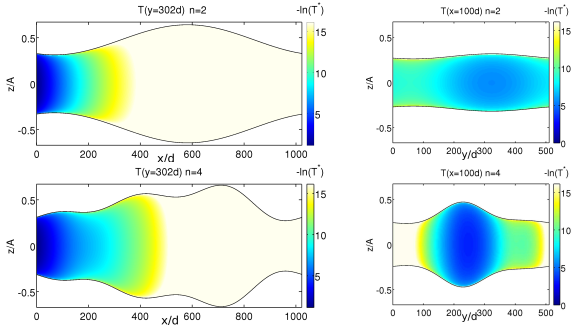

**Figure 12.** Cross section of the scaled temperature in the $n = 2$ (top) and $n = 4$ (bottom) filtered aperture, (left) along the flow, at $y = 302d$ and (right) perpendicular to the flow, at $x = 100d$.

These values are maximal values, and the characteristic amplitude of the relative error can be computed by evaluating $\sigma[e_n]$, the root mean square of these relative errors over each group of values of $\sigma/A$: the dispersion of the relative errors on hydraulic apertures, $\sigma[e(H_n)] = \sqrt{\langle [(H_n - H_{\text{ref}})/H_{\text{ref}}]^2\rangle}$ and on thermal ones, $\sigma[e(\Gamma_n)] = \sqrt{\langle [(\Gamma_n - \Gamma_{\text{ref}})/\Gamma_{\text{ref}}]^2\rangle}$, are displayed as functions of $n$ for each of the five groups on Fig. 14. The error dispersion logically increases as a function of the roughness amplitude $\sigma/A$, and fast decreases as a function of the filtering radius. The largest value of $\sigma[e_n]$, obtained for the group $\sigma/A = 0.5$, drops below 4% on the hydraulic aperture for $n \geq 3$, and on the thermal aperture for $n \geq 4$.

One can directly compute analytically the error made on local values of the aperture $a(x, y)$ by the lowpass filtering procedure, and show that it presents a power law of $n$ as an upper bound, $E(n) \sim n^{-\zeta}$, where $\zeta$ is the Hurst exponent of the aperture field. Technically, the rate of convergence of the Fourier series of the aperture field $a(x, y)$ is determined as follows, from its Fourier transform

$$a(x, y) = \sum_k \tilde{a}(k_x, k_y) e^{-2i\pi(k_x x + k_y y)}. \qquad (9)$$

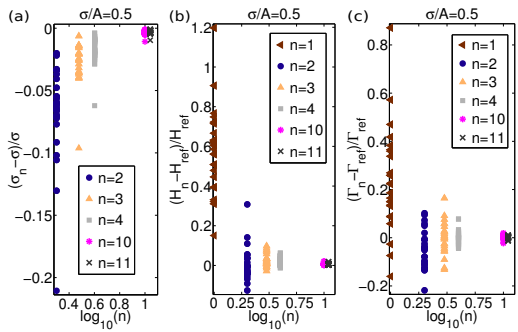

**Figure 13.** Relative error on: (a) the aperture fluctuation $\sigma_n$, (b) the hydraulic aperture, and (c) the thermal aperture, as a function of the degree of the filter, for large aperture fluctuations ($\sigma/A = 0.5$). Cloud of values obtained over 50 independent realizations.

In this series, the error done on the aperture by omitting the terms of the lowpass filter of order $n$, defined by Eq. (8), is

$$E(n) = \sum_{k > n.2\pi/L} \tilde{a}(k, \theta) e^{-2i\pi(kr \cos(\theta - \theta'))}, \qquad (10)$$

where $(k, \theta)$ and $(r, \theta')$ are the radial representations of the vectors $(k_x, k_y)$ and $(x, y)$. The self-affinity of $a$ implies that for $k \neq 0$, its two dimensional Fourier transform scales as $\tilde{a}(k, \theta) \sim C k^{-1-\zeta}$. Hence, for large enough $n$, the sum of the modes comprising the error can be upper bounded as follows, and evaluated using a continuous integral:

$$\begin{aligned}
E(n)^2 &\leq \sum_{k > n.2\pi/L} \|\tilde{a}(k, \theta)\|^2 & (11) \\
&\simeq \sum_{k > n.2\pi/L} C^2 k^{-2-2\zeta} & (12) \\
&\simeq [C^2 L^2/(4\pi^2)] \int_{n.2\pi/L}^{+\infty} k^{-2-2\zeta} 2\pi k dk & (13) \\
&\simeq [C^2 L^2/(2\pi(-1-2\zeta))](2\pi/L)^{-2\zeta} n^{-2\zeta}, & (14)
\end{aligned}$$

which shows that the convergence of the n-lowpass filtered aperture behaves as $E(n) \sim n^{-\zeta}$.

Given this power law dependence of the error on $n$ for the local geometric aperture, it is plausible that the errors done on the hydraulic and thermal apertures present similar power law behavior on $n$, the filtering radius. This hypothesis is consistent with the data, as shown on Fig. 14(c) and (d): the bilogarithmic representation of the dispersion of these errors as a function of the filtering radius is reasonably linear, with a central value around 2 for the slope of the linear regressions (exponent) corresponding to $\sigma[e_n] \sim n^{-2}$ for both errors on hydraulic and thermal apertures – $\sigma[e(H_n)]$ and $\sigma[e(\Gamma_n)]$, defined above. The determined power laws with exponent 2 and prefactors obtained by linear regression in these two figures are also displayed on Fig. 14(a) and (b). These simulations have been performed on elongated fractures of in-plane aspect ratio $J = L_x/L_y = 2$, where $x$ is along the average flow direction and $y$ transverse to it ($z$ being the thickness). It has been shown that the hydraulic and thermal properties of a fracture potentially depend on this aspect ratio (Neuville et al. 2010c). The same analysis as above has also been performed on fractures of aspect ratio $J = L_x/L_y = 1$ (square fractures) and 0.5 (large ones). The results are qualitatively similar to those obtained for the elongated fractures. Up to roughness amplitudes $\sigma/A = 0.4$, the error dispersion of the filter-based approximation drops below 3% for $n \geq 2$ for the hydraulic aperture, and below 4% for the thermal aperture for $n \geq 3$ – see Fig. 14(c), for the case of $\sigma/A = 0.4$ (other values behave similarly, with smaller errors). The error dispersions also behave reasonably as power laws of the filtering radius, $\sigma[e_n] \sim n^{-2}$, as shows the comparison with the fits displayed as continuous lines.

Eventually, the quality of the approximation done by reducing the aperture to the largest Fourier modes can be tested on the following example: similarly to the work done in Neuville et al. (2010c), one can look at the deviation in hydraulic and thermal apertures from a parallel plate model, as a function of the normalized roughness amplitude (for elongated fractures with $J = 2$). The average and standard deviation of $H/A$ and $\Gamma/A$ as functions of $\sigma/A$ are evaluated from 300 independent simulations, each being approximated by its filtered aperture with $n = 4$ radius. This is compared to the results with the full scale description of the aperture for $20\,10^3$ independent simulations (Fig. 15). This figure shows that the full-resolution dependence is not distinguishable for this feature from the results obtained in the filtered aperture (black circles).

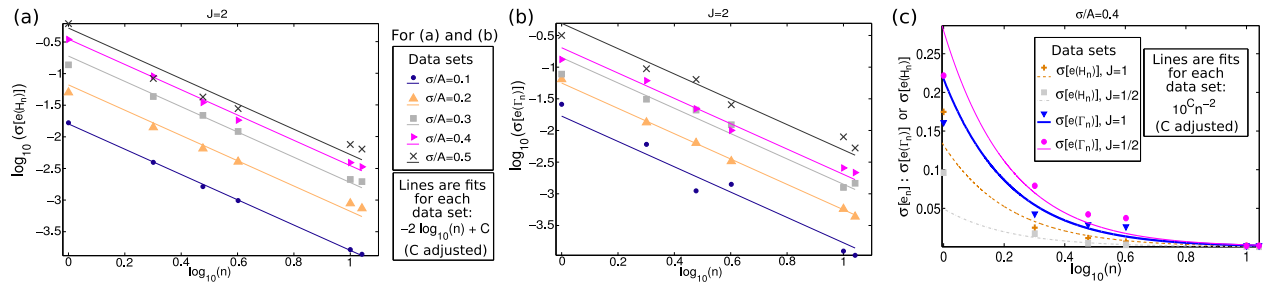

**Figure 14.** Dispersion of the relative errors, $\sigma[e_n]$ (for $n \in \{1-4, 10, 11\}$) on (a) the hydraulic aperture, (b) the thermal aperture, as a function of the degree of the filter, for different aperture fluctuations from $\sigma/A = 0.1$ to 0.5, in bilogarithmic representation, with aspect ratio $J = 2$. Straight lines correspond to a power law behavior $\sigma[e_n] \sim n^{-2}$, corresponding to the central value of linear regressions over each dataset in bilogarithmic space. On (c), the same is shown in bilinear representation for $J = L_x/L_y = 1$ or $1/2$, and aperture fluctuations $\sigma/A = 0.4$.

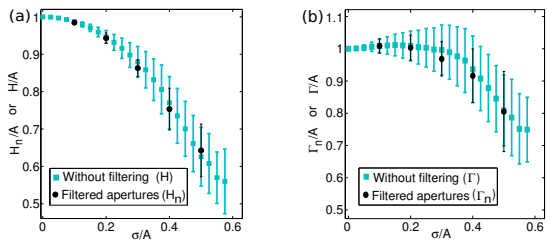

**Figure 15.** Statistical results over numerous synthetic apertures in simulations (with $J = 2$): (a) hydraulic aperture (average and error bars given by plus or minus the root mean square) and (b) thermal aperture, as functions of the roughness amplitude: comparison between the $n = 4$ filter (with 300 aperture field realizations) and the full scale description of the aperture (with $20\,10^3$ aperture field realizations).

## 6  DISCUSSION

Many length scales and time scales intervene when considering the hydraulic and heat transport into a self-affine fracture. Here, the self-affine property of the aperture is supposed to hold for a range of scales, up to the size of the fracture, which is higher than the cutoff length scale introduced in our frequency lowpass Fourier truncation. In practice, the self-affinity property stops at large scales either for geometrical reasons (Neuville et al. 2011a), or for instance if the facing topographies are decorrelated at small scales and correlated at large scales (Méheust & Schmit-tbuhl 2003; Plouraboué et al. 1995, and references therein). In relation to the Stokes flow approximation, the surface is considered smooth below a certain cutoff. If the aperture is also self-affine below this cutoff, this small scale roughness is neglected. The exact role of these small details is studied e.g. by Skjetne et al. (1999). The detailed study of this in terms of heat exchange is a current track for future work under development (e.g. using lattice Boltzmann methods – Neuville, A. 2010a). In the systems we consider, the cutoff of the Fourier filter is much higher than the mean fracture aperture, and also higher than the characteristic thermalization distance (which is proportional to the velocity). For such systems, the Péclet number, defined by $vL_x/\chi$ is high enough so that in-plane advective heat flux is dominant compared to the in-plane diffusive process. It also means that the typical length scale in the fracture plane below which diffusive transport dominates over the advective one, $\chi/v$, is much smaller than the Fourier filter cutoff.

The rock temperature is also supposed to be constant and uniform. This is true for very short time scales when the regime is transitory, or for longer scales at quasi-stationary regimes. Indeed, as the thermal diffusivity of the rock is higher than the fluid one, the fluid temperature evolves faster than that of the rock.

Eventually, the current simulations have all been performed with self-affine aperture fields characterized by a Hurst exponent $\zeta = 0.8$, commonly observed on rock discontinuities. On some joints or rock types, this Hurst exponent changes: e.g., $\zeta = 0.5$ was measured for natural sandstone surfaces (Méheust & Schmittbuhl 2003; Ponson et al. 2007). To generalize the current study to other self-affine joints, a natural question is the impact of this Hurst exponent on the quality of the proposed filtering approach. Performing the whole statistical study carried out here for many different Hurst exponents would be out of the scope of this manuscript, but a hint on this question can be obtained from the convergence of the lowpass filter of the Fourier series describing the aperture field $a$. As was shown earlier, for a given roughness amplitude $\sigma$, the error on the local aperture made by a lowpass filter of order $n$ decays faster than $n^{-\zeta}$. Thus, the convergence of the filtered aperture towards the real aperture field is faster for larger Hurst exponents. The computation of mass and heat flux in better approximated aperture fields is expected to capture better the transport features: the larger the Hurst exponent, the better the quality of the proposed filter-based approximation should be.

## 7 CONCLUSION

Rough aperture fields with a Hurst exponent of $\zeta = 0.8$ correspond to often encountered empty fracture geometries in natural rocks. For such joints, channeling effects affect the flow, and lead to nontrivial hydraulic properties (ratio of the flow over the macroscopic pressure gradient) and efficiency of the thermal exchange between a flowing fluid and the surrounding: these are different from the properties that would correspond to straight channels with flat sides with identical average geometric aperture. A key question, for a given fracture, is what parameters should be kept to model these properties with a sufficient accuracy, while retaining the smallest number of free parameters. We have shown here that the largest scales of the aperture control these properties.

Retaining only the variability at the largest scales (4 modes on top of the average) for example, allows us to reach already an accuracy of 9% over the hydraulic and thermal aperture. Operationally, applying a Fourier filtering of the aperture keeping only the largest Fourier length scales, and solving for the hydraulic and thermal transport in such smoothed aperture fields, provides an excellent approximation to compute the hydraulic and thermal aperture of a given aperture field. If a measurement method determines these few large scale modes, a good deterministic model of these thermal and mass transport properties can be obtained.

## ACKNOWLEDGMENTS

We wish to thank H. Auradou, Y. Méheust, F. Cornet, D. Amitrano, A. Genter, E. Flekkøy, A. Cochard and P. Baud for fruitfull discussions. We also acknowledge the financial support of the ANR ECCO, ANR TRIGGERLAND, INSU, and of the Alsace region through the REALISE program.